\documentstyle[12pt,aaspp4]{article}

\begin{document}

\def\kms{$\rm {km}~\rm s^{-1}$}
\def\ni{\noindent}
\def\Msun{{\rm M}_\odot}
\def\deg{\ifmmode^\circ\else$^\circ$\fi}
\def\etal{{\it et~al.}}

\title{Fabry-Perot Observations of Globular Clusters III: M15}
\author{Karl Gebhardt\altaffilmark{1}} 
\affil{Dept. of Astronomy,Dennison Bldg., Univ. of Michigan, Ann Arbor
48109}
\affil{gebhardt@astro.lsa.umich.edu} 
\author{Carlton Pryor\altaffilmark{1}, T.B. Williams\altaffilmark{1}}
\affil{Department of Physics and Astronomy, Rutgers, The State
University of New Jersey, P.O. Box 849 Piscataway, NJ 08855-0849}
\affil{pryor@physics.rutgers.edu,williams@physics.rutgers.edu}
\author{James E. Hesser\altaffilmark{1} and Peter B. Stetson}
\affil{Dominion Astrophysical Observatory, Herzberg Institute of
Astrophysics, National Research Council of Canada}
\affil{5071 W. Saanich Road, R.R.5, Victoria, B.C., V8X 4M6, Canada}
\affil{James.Hesser@hia.nrc.ca,Peter.Stetson@hia.nrc.ca}
\altaffiltext{1}{Visiting Astronomer, Canada-France-Hawaii Telescope,
operated by the National Research Council of Canada, the Centre
National de la Recherche Scientifique of France, and the University of
Hawaii.}

\begin{abstract}

We have used an Imaging Fabry-Perot Spectrophotometer with the
Sub-arcsecond Imaging Spectrograph on the Canada-France-Hawaii
Telescope to measure velocities for 1534 stars in the globular cluster
M15 (NGC~7078) with uncertainties between 0.5 and 10 \kms.  Combined
with previous velocity samples, the total number of stars with
measured velocities in M15 is 1597.  An average seeing of 0.8\arcsec\
allowed us to obtain velocities for 144 stars within 10\arcsec\ of the
center of M15, including 12 stars within 2\arcsec. 

The velocity dispersion profile for M15 remains flat at a value of
11~\kms\ from a radius of 0.4\arcmin\ into our innermost reliable
point at 0.02\arcmin\ (0.06 pc).  Assuming an isotropic velocity
dispersion tensor, this profile and the previously-published surface
brightness profile can be equally well represented either by a stellar
population whose M/L varies with radius from 1.7 in solar units at
large radii to 3 in the central region, or by a population with a
constant M/L of 1.7 and a central black hole of 1000~$\Msun$.  A
non-parametric mass model that assumes no black hole, no rotation, and
isotropy constrains the mass density of M15 to better than 30\% at a
radius of 0.07 parsecs.  The mass-density profile of this model is
well represented by a power law with an exponent of --2.2, the value
predicted by models of cluster core-collapse.  Using the assumption of
local thermodynamic equilibrium, we estimate the present-day mass
function and infer a significant number of 0.6--0.7~$\Msun$ objects in
the central few parsecs, 85\% of which may be in the form of stellar
remnants.

Not only do we detect rotation; we find that the position angle of the
projected rotation axis in the central 10\arcsec\ is 100\deg\
different from that of the whole sample.  We also detect an increase
in the amplitude of the rotation at small radii. Although this
increase needs to be confirmed with better-seeing data, it may be the
result of a central mass concentration.

\end{abstract}


\section{INTRODUCTION}

Imaging Fabry-Perot (FP) spectrophotometers have become powerful tools
for studying the dynamics of the central regions of globular clusters.
The samples of individual stellar velocities that have been obtained
with a FP are some of the largest available for globular clusters
(Gebhardt \etal\ 1994, 1995, hereafter Paper~1 and Paper~2,
respectively). In addition, FP observations of the integrated light at
the center of a cluster provide a two-dimensional velocity map of
that region with unprecedented detail.  With the dramatic increase in
the quality of the kinematic data available for globular clusters,
more powerful analysis techniques become possible.  Non-parametric
techniques can provide an unbiased estimate of the dynamical state of
a cluster (Merritt 1993a,b), such as the rotation and velocity
dispersion properties (Paper~1 and 2).  Combining the dispersion and
surface brightness profiles with the Jeans equation directly yields
the mass density profile and the mass function (Gebhardt \& Fischer
1994, hereafter GF, Merritt \& Tremblay 1994).  The distribution of
mass determined by this technique can be used to test the predictions
of Fokker-Planck and N-body simulations.

As one of the densest nearby objects known, M15 has significant
implications for the formation and evolution of dense stellar systems.
It has, therefore, been the target of intense observations and
theoretical modeling.  Images taken with the refurbished Hubble Space
Telescope (Guhathakurta \etal\ 1996) have shown that the luminosity
density may rise all of the way into the center and do not support the
core with a radius of about 2\arcsec\ previously proposed by Lauer
\etal\ (1991).  Measurements of the dynamical state of M15 have also
produced controversy.  Is there a central cusp in the dispersion
profile (Peterson, Seitzer, \& Cudworth 1989), or does the profile
remain flat near the center at 11 \kms\ (Paper~1, Dubath \& Meylan
1994, Dull \etal\ 1996)?  Somewhat conflicting amounts of rotation
have been measured near the center by Peterson (1993) and Paper~1.

GF derived a non-parametric mass model for M15 using the velocities
from Paper~1, but this has not been compared with the results of
either Fokker-Planck (Grabhorn \etal\ 1992, Dull \etal\ 1996, Einsel
\& Spurzem 1996) or N-body (Makino \& Aarseth 1992, McMillan \&
Aarseth 1993) evolutionary models.  The strongest constraints on the
models come from comparing them with the kinematics in the central few
arcseconds, since this is where the effects of mass segregation, core
collapse, and a possible central massive black hole are the most
pronounced.  Unfortunately, the uncertainties in and disagreements
between the observations in this region have made comparisons
difficult.

To improve our knowledge and the significance of the rotation, the
value of the dispersion within 2\arcsec\ of the center of M15 requires
observations with higher angular resolution.  The tools of
crowded-field photometry can be directly applied to the
two-dimensional data from a FP, making them easier to interpret for
dense stellar systems than slit spectra and yielding the largest
possible number of stellar velocities.  Using a FP on telescopes and
with instruments that provide the best seeing is a promising approach
to resolving the mystery of the center of M15.  This paper is the
first reporting results from our study of globular cluster dynamics
using data taken at the Canada-France-Hawaii Telescope (CFHT). These
data have better seeing than that in Papers~1 and 2, which were taken
at CTIO.

The rest of this paper is organized as follows.  Sec.~2 discusses the
Fabry-Perot observations and the reductions.  Sec.~3 describes
measurements of the first and second moments of the velocity
distribution.  Sec.~4 presents the mass modeling and the mass function
estimates.  Sec.~5 discusses the results.

\section{THE DATA}

\subsection{Instrumentation}

We used an imaging Fabry-Perot spectrophotometer with the
Sub-arcsecond Imaging Spectrograph (SIS) at the CFHT on May 20--23,
1994, and May 12--17, 1995.  The optical design of SIS is described in
Morbey (1992) and Le Fevre \etal\ (1994).  Light is fed to the
spectrograph by a tip-tilt mirror which is driven at about 20~Hz using
the signal from a quadrant detector at the focal plane in order to
compensate for image motion.  We placed the Rutgers narrow etalon,
which has a spectral resolution of 0.8~\AA\ FWHM at 6500~\AA, in the
collimated beam.  This etalon is fully compatible with the CFHT etalon
controller and its coatings are suitable for work at the H$\alpha$
line.  The order-selecting filter was in the collimated beam below the
etalon and was tilted to eliminate ghost images due to filter
reflections.  Before being tilted, the filter had a 16\AA\ FWHM
centered at 6569\AA\ and a peak transmission of 83\%.  It was borrowed
from the Dominion Astrophysical Observatory.

The SIS+FP setup provides a $3\arcmin\times 3\arcmin$ field of view.
The front-side illuminated Loral3 CCD imaged this field, and we binned
$2\times 2$ during readout to obtain 1024x1024 pixels at a scale of
0.173\arcsec\ per pixel.  The SIS guide probe is mounted on a
45\arcsec\ wide arm that runs across the entire field.  Fortunately,
the probe's field of view is $4\arcmin\times 3\arcmin$; with the rich
star fields of globular clusters, we were able to choose V~=~13~--~14
guide stars that resulted in little or no vignetting of our images by
the probe.  Sampling the position of the star at 40~Hz seemed adequate
to follow the image motion seen on a real-time display.  Fast guiding
produced an approximately 15\% improvement in the FWHM of the images.
Thus, we were able to reduce the area of star images by about 30\%,
which significantly increased the number of stars for which we could
obtain velocities.

\subsection{Observations and Data Reduction}

The FP observing and reduction procedures were the same as those
described in Paper~1.  We took a series of exposures stepped by
0.33~\AA\ across the H$\alpha$ absorption line.  Projector flats were
obtained in the evening or morning for every wavelength setting of the
etalon used that night.  Approximately hourly exposures of a deuterium
lamp, which provided both H$\alpha$ and D$\alpha$ emission lines,
monitored the wavelength zero-point and provided the primary
wavelength calibration.  This calibration was supplemented by
exposures of several neon lines taken throughout the run.  A small
offset between the focal planes of the guide probe and the SIS
collimator resulted in reflections from the etalon being so out of
focus that they were not detectable.  We thus made no corrections for
reflected light.

We obtained 17 15-minute exposures of M15 in 1994 and 15 such
exposures in 1995. Frame-to-frame normalizations for non-photometric
conditions, determined as described in Paper~1, were as large as a
factor of two for the 1994 data. For the 1995 data, the normalizations
only varied by 20\%.  The average FWHM of the stellar images was
0.8\arcsec\ (the 1995 run) and 1.1\arcsec\ (1994). The overall
throughput of the system (telescope+FP+CCD) was about 15\%, and for a
V=16 star we were able to obtain a signal-to-noise of 18 in a
15~minute exposure under photometric conditions.

A significant change from our previous reduction procedure was
employing HST images directly to assist in the photometry of the
crowded central regions of M15.  This was made possible by using
ALLFRAME (Stetson 1994), which measures positions and brightnesses for
all of the stars in a set of images simultaneously.  Coordinate
transformations with six coefficients map a location in one frame to
the corresponding location in another.  ALLFRAME makes maximum use of
the positional information present in the set of frames, improving the
photometry in every frame.  Our initial ALLFRAME reduction used HST
frames from Guhathakurta~\etal\ (1996) and all of the frames from the
1994 and 1995 CFHT runs.  Unfortunately, the line profiles resulting
from this photometry were very noisy.  We believe that this resulted
from faint stars which are isolated in the HST images, but very
blended with a brighter neighbor in the CFHT images.  The partitioning
of light among the crowded stars varied wildly from CFHT frame to CFHT
frame.  The ``sky'' level determined in the CFHT frames also tended to
be incorrect.  These problems are clearly related to the large
difference in resolution between the CFHT and HST images.

A different initial list of stars to be reduced or some other change
of procedure might have solved these problems.  However, what we chose
to do was to photometer each CFHT frame individually using ALLSTAR,
but holding the positions of the stars fixed at those found from the
ALLFRAME reduction.  This allowed ALLSTAR to determine, based on a
uniform chi-square criterion, whether there was adequate information
in an individual frame to determine a star's magnitude with acceptable
precision.  The line profiles produced by this procedure were
smoother, but crowding probably still introduces some additional
uncertainties into our velocities inside a radius of about 0.3\arcmin.
This is discussed in somewhat more detail below.

We also used the HST photometry to determine which stars in the CFHT
frames were sufficiently contamination by light from neighbors that
their measured velocity could be affected.  The criterion is based on
Monte Carlo simulations and is the same as used in Paper~1.  Due to
the 0.8\arcsec\ FWHM seeing of the CFHT SIS frames, only eight stars
were excluded from the final sample.

\subsection{Velocity Zero-Point and Uncertainties}

Radial velocities for M15 stars were previously obtained by Peterson
\etal\ (1989, hereafter PSC), who used the echelle spectrograph and
intensified Reticon detector on the MMT, and by us (Paper~1). The
present study includes 66 stars in common with PSC and 188 in common
with Paper~1.  Based on this overlap, both the 1994 and 1995 raw
velocities are $7.4\pm 0.4$~\kms\ more negative than those of Paper~1,
which were adjusted to the zero-point of PSC.

While a zero-point offset between radial velocities measured with
different instruments is not surprising, this is the largest offset
that we have seen for FP data.  In order to investigate this further,
we observed the radial velocity standards HD107328 and HD182572 during
the 1995 run.  These data were taken and reduced in a fashion very
similar to the globular cluster stars.  With only one star in the
field we could not derive transparency normalizations for the
individual frames, but conditions were photometric and the exposure
times were only 15~s.  Observations of HD107328 with the star near the
center of the field on one night and 370~pixels from the center on
another yielded velocities of $29.85\pm 0.03$~\kms\ and $30.35\pm
0.03$~\kms, respectively.  An observation of HD182572 on a third night
with the star near the center of the field produced a velocity of
$-106.79\pm 0.05$~\kms.  The quoted uncertainties are simply the
uncertainties in the fitted line centroids, and may be underestimated
since we do not include noise from possible transparency fluctuations.

The standard velocity for HD107328 is 36.6~\kms\ and for HD182572 is
$-100.3$~\kms\ (Latham \& Stefanik 1991).  The differences between the
three raw FP velocities and the standard velocities are $-6.75$, $-6.49$,
and $-6.25$~\kms.  The mean difference is $-6.5$~\kms, with an RMS
scatter around this mean of 0.25~\kms.  This zero-point offset is
probably caused by some combination of small wavelength, flatfielding,
and normalization errors and a difference between the centroid of the
stellar H$\alpha$ line measured by our line-fitting program and the
tabulated laboratory wavelength of H$\alpha$.  The scatter between the
three velocity differences is satisfyingly small and argues that,
whatever its cause, the velocity offset will not significantly increase
the scatter within our cluster sample.  The PSC velocity zero-point was
based on spectra of the twilight sky, so we consider the agreement
between the zero-points derived from the standards and from the M15
stars to be acceptable.  We adopt the latter for the remainder of this
paper.

Figure 1 plots V$_{\rm Paper 1}$--V$_{\rm FP95}$ vs. V$_{\rm Paper
1}$.  The comparison with PSC is similar and not shown here.  Several
binary star candidates are apparent in Fig.~1, but discussion of these
is deferred to a future paper.  The RMS difference measured with a
bi-weight estimator, which is insensitive to a few outliers, is
5.0~\kms\ after correcting the zero-point of the 1995 data.  This
implies a typical velocity uncertainty of 3.5~\kms.  Comparing this to
the average internal velocity uncertainty calculated from fitting the
the H$\alpha$ line profile suggests that we need to add 1~\kms\ in
quadrature to these internal uncertainties.  This additional
uncertainty could be due to the fits to the line profiles
underestimating the uncertainties or to the velocity ``jitter'' of
luminous cluster giants (Gunn \& Griffin 1979, Mayor \etal\ 1983).

There are 461 stars with velocities from both the 1994 and 1995 data
and this provides a large, homogeneous sample with which to explore
the FP measurement uncertainties.  We determined what value, added in
quadrature to the internal uncertainties, produced the most uniform
distribution of chi-square probabilities.  Probabilities smaller than
5\% are excluded to remove the effect of real velocity variables.
This method was first employed by Duquennoy \& Mayor (1991) and the
approach that we used to judge uniformity is described in
Armandroff~\etal\ (1995).  The best additive uncertainty was zero for
the entire sample, though values as large as 1.0~\kms\ are acceptable.
About 10\% of the stars have probabilities under a few percent, which
would be a surprisingly large fraction of real variables.  The stars
with smaller average measurement uncertainties had more low
probabilities, suggesting that the brighter stars should have larger
additive uncertainties.  Limiting the sample to the 118 stars with
approximate $V$ magnitudes (described in the next section) less than
15 produced an additive uncertainty of 1.4~\kms.

This bright subsample also has about 10\% of the stars with
probabilities below a few percent and we noticed that a larger than
expected number of these were at small radii.  Focusing on the 393
stars outside of a radius of 0.4\arcmin, the most uniform probability
distribution resulted from an additive uncertainty of 0.4~\kms\ for
stars with any magnitude and from a value of 1.1~\kms\ for the 91
stars brighter than magnitude 15.  This difference very likely
reflects the well-known velocity jitter of luminous cluster giants and
suggests that much of the additional uncertainty reflected in the
comparisons with the Paper~1 and PSC velocities, which are mostly for
brighter stars, comes from this source.

For the 68 stars with both 1994 and 1995 velocities inside a radius of
0.4\arcmin, the most uniform probability distribution results for an
additional uncertainty of 3.2~\kms.  With this value, there is no
evidence for an excess of stars with low probabilities.  Most of these
stars are brighter than a magnitude of 15.8 because of crowding, but
it is still clear that the FP velocities suffer from additional
uncertainty at small radii.  Errors in the stellar photometry induced
by crowding are almost certainly the cause.  The line profiles at
small radii show more scatter around the fitted profiles than do those
at large radii.  The PSC stars in this region are probably the
brightest and least crowded.  Comparing the PSC and 1995 velocities
for the 26 stars in this region required an additional uncertainty for
the CFHT velocities of only 0.7~\kms.

All of the additional uncertainties described above are much smaller
than the cluster dispersion and, thus, have little effect on the
dynamical analyses.  We have thus adopted the simple prescription of
adding 1~\kms\ in quadrature to the uncertainties derived from the
line profile fits for the 1994 and 1995 velocities.  We have performed
our dynamical analyses using an additional uncertainty of 3~\kms\
instead of 1~\kms and find the same results.  We have also performed
the analyses both including and excluding those stars with small
chi-square probabilities and find no difference in the results.  The
results presented below include these stars.

\section{RESULTS}

\subsection{The Individual Stellar Velocities}

We have combined five sets of M15 velocities: the PSC measurements and
Fabry-Perot measurements from 1991 (Paper~1), 1992 (Paper~1), 1994
(this paper), and 1995 (this paper).  Each star for which there was
any ambiguity in the matching among datasets (either in the position
or in the velocities) was examined carefully to ascertain the accuracy
of its identification.  Table~1 presents the mean stellar velocity for
every measured M15 star.  Col.~1 is the star ID; Cols.~2 and 3 are the
x and y offset from the cluster center in arcseconds (measured
eastward and northward, respectively, from the center given in
Guhathakurta \etal\ 1996); Cols.~4 and 5 present the mean velocity and
its uncertainty; Col.~6 gives the FP magnitude (estimated from the
fitted continuum of our 3~\AA\ of spectrum); and Col.~7 lists either
the probability of the chi-square from the multiple measurements
exceeding the observed value -- a low probability suggests a variable
velocity, though see \S~2.3 -- or a note which is explained below.
The average velocity given was calculated weighting by the measurement
variances, and the inverse of the uncertainty is the square root of
the average of the individual inverse variances.  The uncertainties
include the 1.0~\kms\ added in quadrature that was discussed in the
previous section.

The wavelength gradient with radius in the FP images (Papers~1 and 2)
causes a bias in the velocity sample due to the limited wavelength
coverage of the FP data.  Stars at large radii and with velocities
sufficiently more positive than the cluster mean will have incomplete
wavelength coverage of the line, which makes the velocity impossible
to measure.  We therefore have to choose a radius from the center of
the FP field where this effect may begin to affect the dynamical
analysis.  The SIS optics at CFHT produce a gradient of only 4~\AA\ as
compared to 6~\AA\ at CTIO.  As a result, we obtain a larger unbiased
field at the CFHT than at CTIO for the same number of frames:
1.2\arcmin\ in radius vs. 1.0\arcmin, respectively.  The stars that
are beyond this radius are not used in the dynamical analysis and have
``bias'' in the final column of Table~1.  Stars determined to be
non-members based on their radial velocity have ``non'' in the final
column of Table~1.

The stars with a low probability (P$<0.01$) in Col.~7 are good
candidates for stars whose velocity varies due to jitter or binary
orbital motion.  However, identifying binary candidates is difficult
because of the complex situation with the velocity uncertainties
discussed in \S~2.3.  We, therefore, postpone discussion of the binary
candidates and the binary frequency to a future paper.

The individual velocity measurements from the 1994 and 1995 CFHT runs
and from the previously published samples are given in Table~2.
Col.~1 is the star ID; Cols.~2 and 3 are the velocity and uncertainty
from the 1995 CFHT data; Cols.~4 and 5 are from the 1994 CFHT data;
Cols.~6 and 7 are from the 1992 CTIO data (Paper~1); Cols.~8 and 9 are
from the 1991 CTIO data (Paper~1); and Cols.~10 and 11 are from PSC.
The uncertainties listed in Table~2 do not contain the additional
uncertainty that is used in Table~1.

Recently, Dull~\etal\ (1996) reported velocities for 132 member stars
in the central 1\arcmin\ of M15. We have repeat velocity measurements
for all of their stars, and our average uncertainty is around
1.5~\kms, while theirs is 4~\kms. Because they do not list the
uncertainties for the individual stars, we are unable to combine their
velocities with our dataset in the same rigorous fashion.  Since our
uncertainties are smaller for their stars and since our dataset
contains over a factor of ten more velocity measurements, the
dynamical analysis will not be affected by not including their
velocities.

Tables~1 and 2 contain only stars with velocity uncertainties smaller
than 10~\kms.  We have determined that it is worthwhile to use stars
with uncertainties as large as this, even though the velocity
dispersion of the cluster is around 11~\kms.  We formed velocity
samples that excluded stars with uncertainties larger than values as
small as 4~\kms.  Dynamical analyses of these samples showed no
significant change in the results other than an increase in the size
of the confidence bands for the measured quantities when fewer stars
were used.  Thus, we employed the complete velocity sample in the
analysis presented in this paper. Only four stars were found to be
non-members due to their large velocity difference from the cluster
mean velocity; they were excluded.  An additional 22 stars were not
used because they are at large radii where the dynamics may be biased
by the limited spectral coverage.  Of the remaining 1575 stars with
uncertainties less than 10~\kms, 1100 have uncertainties less than
5~\kms and 685 less than 3~\kms. Inside of 0.1\arcmin\ we were able to
measure velocities for 80\% (71 of 89) of the stars brighter than
V=18.

Figure 2 plots the individual stellar velocities and their
uncertainties against the distance from the center of M15. This figure
suggests that the local mean velocity for M15 becomes more positive in
the central 0.1\arcmin.  The weighted mean velocity of the 71
innermost stars is $-105.6\pm 1.3$~\kms, while the mean velocity of
the whole sample is $-107.8\pm 0.3$~\kms.  Therefore, the apparent
change in velocity is significant at the 2$\sigma$ level. We have
looked for calibration and/or normalization errors and are confident
that these cannot produce such a large velocity shift.  Comparison
with previous authors also increases our confidence that our velocity
measurements are not biased.  Dubath \& Meylan (1995) have measured
velocities for 14 stars in the central 0.1\arcmin\ of M15.  The mean
velocity for these 14 stars is $-103\pm 4$~\kms, consistent with our
measurement. We feel that the variation of the mean velocity with
radius is simply due to statistical noise.

\subsection{Rotation}

We detect rotation in M15.  Using maximum likelihood techniques to fit
a sinusoid to the entire sample of individual stellar velocities as a
function of their position angle yields a rotation amplitude of
$2.1\pm 0.4$~\kms.  The position angle of the maximum positive
rotation velocity, measured from north through east, is $197\arcdeg\pm
10\arcdeg$.  Both of these values agree with our previous measurement
of $1.4\pm 0.8$~\kms\ and $205\arcdeg \pm 33\arcdeg$ (Paper~1).  Monte
Carlo simulations show that a 2.1~\kms\ rotation amplitude is expected
by chance when no rotation is present less than 0.1\% of the time.

We can study the two-dimensional structure of the projected rotation
by using a thin-plate, smoothing spline (Wahba 1980, Bates \etal\
1986) to produce a map of the mean velocity as a function of location
on the sky. This method allows us to check for twisting position
angles and other unusual properties which might otherwise be missed
using a one-dimensional representation of the data. The smoothing
spline is fit to the velocity data with the smoothing length chosen by
generalized cross-validation (GCV).\footnote{GCV uses a jackknife
approach. For a given smoothing, one point is removed from the sample
and the spline derived from the remaining $N-1$ points is used to
estimate the value for the removed point. This procedure is repeated
for each point, and the optimal smoothing is that which minimizes the
sum of squared differences between the actual data points and the
estimated points. A good explanation of GCV can be found in Craven \&
Wahba (1979) and Wahba (1990) (see Gebhardt \etal\ 1996 for an
application to a one dimensional problem).}  Figure~3 shows the
resulting isovelocity contours for M15. The points represent the
positions of the stars that have a measured velocity.  The spacing of
the contours in Fig.~3 is 1~\kms; given that the cluster dispersion is
about 11~\kms, the estimate of the two-dimensional velocity map may be
noisy, despite its large number of radial velocities.  The contours
do, however, suggest a possible twisting of the rotation axis position
angle (PA) with radius, as we discuss below.

To get a higher S/N estimate of the two-dimensional velocity structure
we will assume axisymmetry.  This assumption is invalid if the PA
twists.  However, we will regard the possible twisting in Fig.~3 as a
perturbation to the global velocity structure and use the PA measured
for the whole sample as the symmetry axis.  We can then reflect
stellar positions about both the rotation axis and a line
perpendicular to that axis and re-estimate the velocity map, a
procedure that will effectively increase the number of data points by
a factor of four.  Figure~4 plots the resulting axisymmetric
isovelocity contours in a single quadrant, calculated using the same
smoothing spine as above.  Here the y and x axes are along and
perpendicular to the rotation axis, respectively.

The projected rotation of a spherical system with solid-body internal
rotation (i.e., rotation velocity constant on cylinders) will have
isovelocity contours parallel to the rotation axis.  This is
guaranteed because, along the line-of-sight at a fixed projected
distance from the axis, the linear decrease in the radial velocity due
to the changing projection of the rotation on the line-of-sight is
offset by the linear increase in the internal solid-body rotation
profile.  The vertical contours of Fig.~4 suggest that the rotation in
M15 is consistent with an internal solid-body rotation profile.  A
similar non-parametric estimate of the symmetrized rotation map for
47~Tuc shows the rotation decreasing below that expected for a
solid-body form beyond a radius of 2\arcmin\ (Paper 2).  The
half-light radius of M15 is 2.9$\times$ smaller than that of 47~Tuc
(Trager \etal\ 1993; primarily reflecting the larger distance of M15),
so the two clusters may have real differences in the form of their
rotation.  More velocities are need at large radii in M15 to be sure
of this, however.

Figure~3 suggests that the rotation PA changes from north-south at
small radii to more east-west at large.  We can check the significance
of this result by radially binning the individual stars and estimating
the rotation properties in each bin.  We use a maximum likelihood
estimator that yields the amplitude and phase for a sinusoidal
variation of the mean velocity with position angle, and the dispersion
around the sinusoid, while holding the mean velocity of the sample
fixed at the cluster mean.  The results are given in Table~3 and
plotted in Fig.~5.  The solid points in the two panels of Fig.~5 are
the rotation amplitude and position angle as a function of radius for
five bins containing 300 stars and a final outer bin of 74 stars.  The
amplitude of the rotation from Fig.~5 will not match that of Fig.~4
since, unlike Fig.~5, Fig.~4 is derived under the assumption of
axisymmetry.  The position angle of the rotation does increase beyond
a radius of 1\arcmin, consistent with the twisting of the isovelocity
contours in Fig.~3.  However, this result depends somewhat on the
binning adopted and more velocities at large radii are needed to
confirm this feature.

The two innermost solid points in Fig.~5 suggest that the position
angle of the rotation also changes at small radii, though the rotation
amplitude of even the larger innermost point is small enough that it
could occur by chance about 13\% of the time.  This feature might not
appear as a twisting of the contours in Fig.~3 because of the
smoothing.

To obtain a more detailed picture of the rotation profile at small
radii, we turn to the procedure of using the integrated light
developed in Paper~1. The Fabry-Perot observations provide
two-dimensional images that yield an integrated-light spectrum at
every sufficiently bright pixel and, hence, the two-dimensional
velocity structure of the cluster. The integrated light provides a
better estimate of the rotation properties than the individual
velocities because it effectively samples a larger number of stars,
which will limit the noise due to the cluster dispersion. We
azimuthally bin and average velocities from the integrated-light
velocity map and find the best-fit sinusoid. These results are also
listed in Table~3 and plotted as open circles in Fig.~5.  The results
from the 1994 and 1995 CFHT data are similar and we show only the
latter since the seeing was better in 1995.

The integrated light map only has adequate S/N within 20\arcsec\ of
the center; beyond that, we have to rely on the stellar velocities to
determine the rotation properties.  In the region where there is
significant overlap, the rotation properties measured by the
integrated light and the stars agree in both amplitude and position
angle.  The integrated-light results presented here are reasonably
consistent with the corresponding results from Paper~1 for the region
within a radius of 15\arcsec\ (with a mean radius of 8\arcsec), which
were an amplitude of $1.7\pm 0.3$~\kms\ and a position angle for the
maximum positive rotation velocity of $220\arcdeg\pm 11\arcdeg$.  Our
new results support a change in the position angle of the rotation
axis at small radii and suggest that the rotation amplitude increases
as well.

However, in the central regions, seeing and sampling noise become an
important consideration for interpreting the integrated-light rotation
properties. Peterson (1993) reported a streaming motion with a full
range of 15~\kms\ in the central 1\arcsec\ of M15. Dubath \& Meylan
(1995), using CORAVEL with slits at various positions in the center of
M15, argued that Peterson's result may have been due to sampling
noise. The two brightest stars in the central region (AC212 and AC215)
have velocities differing by 30~\kms\ and are aligned along the
direction for which the streaming motion was reported.  Using our
two-dimensional velocity map, we find a result similar to that of
Dubath \& Meylan. By synthesizing the slit position and size of
Peterson's observation, our velocity map shows the same change in the
mean velocity with position as Peterson measured; however, the result
is due to contamination from the wings of the two bright stars,
emphasizing the importance of understanding the sampling noise
(Dubath~1993). Our integrated-light rotation measurements will suffer
from the same problem.  For seeing of 0.8\arcsec, we estimate that the
radius at which two stars may cause a rotation signature is around
2\arcsec. The vertical dashed line in Fig.~5 is at this
radius. However, the PA changes smoothly from larger radii to the
smallest radius, giving us confidence that the increase in the
rotation amplitude at small radii seen in Fig.~5 is a physical effect,
and not due to sampling.

We can further check the significance of the central increase in the
rotation by exploiting the two-dimensional data of the FP. We subtract
the five brightest stars in the central 4\arcsec, re-calculate the
velocity map, and estimate the rotation from the map as before. This
procedure may show the influence of the bright stars on the inferred
central rotation. The rotation obtained in this way does not show the
characteristics seen in Fig.~5; in fact, there is no detectable
rotation when the five bright stars are subtracted. However, this
result may be due mainly to noise in the subtraction. Inspection of
the line profiles for the integrated light before and after
subtraction demonstrates that the subtraction added significant noise,
making determination of the line centroid uncertain. Another method,
instead of subtracting the bright stars, is to ignore the pixels out
to some radius underlying those stars. This procedure gives similar
results to those in Fig.~5, but it is questionable since it depends on
the radius chosen. Yet another check is to measure the rotation from
the stellar velocities inside of 2\arcsec. For the 12 stars inside of
2\arcsec, we measure a rotation amplitude of $8.5\pm3.8$~\kms\ and a
PA of $303$\deg$\pm26$, consistent with the integrated light
profile. We conclude that the rotation inside of 2\arcsec\ appears
significant. However, better-seeing data are needed to reduce the
sampling uncertainties and we, therefore, have chosen a radius of
2\arcsec\ as the boundary inside of which our results may have been
affected by this.

\subsection{Velocity Dispersion}

Fig.~6 plots for our M15 sample the absolute magnitude of the
deviation of each stellar velocity from the cluster mean velocity
vs. radius.  The points are the velocity measurements, and the solid
and dashed lines are the LOWESS estimates of both the velocity
dispersion and the 90\% confidence band, respectively (see Paper~2 for
details). The dispersion does not increase significantly inside a
radius of 0.4\arcmin.

This dispersion profile is consistent not only with that of Paper~1
but also, outside of a radius of 1\arcsec, with the profile increasing
towards smaller radii which was found by PSC. However, in the central
1\arcsec, PSC reported a cusp in the velocity dispersion based on
integrated-light measurements. Our larger sample, which has four stars
within 1\arcsec\ of the center, shows no evidence for a central cusp.
PSC's high velocity dispersion measurement may have been due to
sampling noise (Dubath 1993, Zaggia~\etal\ 1993). Our estimate of the
dispersion uses a smoothing length, which prevents our estimated
profile from changing quickly. However, the dispersion of the four
stars in the central 1\arcsec\ is 11.6 \kms, consistent with the
LOWESS estimate of Fig.~6; this estimate is also consistent with the
results of Dubath \& Meylan (1994) and Dull~\etal\ (1996).

\section{MASS MODELING}

\subsection{Non-Parametric Estimate}

We have used the non-parametric mass modeling technique of GF to
estimate the mass density profile for M15.  Under the assumptions of
spherical symmetry and an isotropic distribution of velocities at all
points, the Abel equations provide estimates of the deprojected
luminosity density and velocity dispersion profiles, given the
projected quantities.  With the same assumptions, the Jeans equation
then yields a unique mass density profile.  We use the velocity
dispersion profiles from Fig.~6, and the surface brightness profiles
from Grabhorn~\etal\ (1992) and Guhathakurta~\etal\ (1996).  Figure~7
plots the resulting non-parametric estimates of the mass density and
the mass-to-light ratio (M/L) profiles.  The solid lines are the
bias-corrected estimates and the dotted lines are the 90\% confidence
bands. The bias is estimated through bootstrap resamplings (see GF).

The dashed line in the upper panel of Fig.~7 has a slope of --2.23,
which is the theoretical prediction for the region outside of the core
in a core-collapse cluster (Cohn 1980). Theory and observation are
clearly in good agreement. Fig.~8 plots the logarithmic slope of the
mass density profile and its 90\% confidence band. Since derivatives
are inherently noisier, the uncertainties on the slope are large;
particularly at large radii the slope is highly uncertain. Not that
the confidence bands depend on the amount of smoothing which has been
introduced in the velocity dispersion estimate. Also, the
uncertainties in the surface brightness profile are not considered
here and may significantly increase the uncertainties in the slope
estimate. The confidence band shown in Fig.~8 only reflects the
uncertainty based on the smoothing used for Fig.~7; a more rigorous
treatment would have to include the surface brightness uncertainties
and try different velocity dispersion smoothings.

Phinney~(1992, 1993) used measurements of pulsar accelerations to
estimate the central density in M15 and found a value of greater than
$2\times10^6~\Msun$/pc$^3$.  We are able to follow M15's mass density
over five decades of density (two of radius). In the central
1.4\arcsec\ (0.07 parsecs) the inferred density reaches $2\times
10^6~\Msun$/pc$^3$ with 1$\sigma$ uncertainties under 30\%, in
concordance with Phinney's value.

The estimated M/L$_{\rm V}$ for M15 increases from about 1.2 at a
radius of 10\arcsec\ (0.5~pc) to nearly 3 in the central 0.05 parsecs.
There is a similar, but more statistically significant, increase in
the outer 8 parsecs. We attribute these increases to a concentration
of heavy stellar remnants in the central regions and of low-mass stars
in the outer parts.

Rotation has not been included in this analysis.  However, the
dispersion about the average rotation velocity of 2.0~\kms\ for our
M15 sample is 9.8~\kms.  The ratio of the squares of these quantities,
which measures the dynamical significance of rotation, is 0.04.  In
regions where the rotation velocity is as large as 4~\kms (see
Fig.~5), this ratio approaches 0.2.  Still, our dispersion values in
Fig.~6 were increased by ignoring the rotation, which approximately
mimics the effect of rotation in the Jeans equations, so our mass
distribution calculated without including rotation will likely be a
good approximation.

GF also presented a mass density profile for M15 and their profile
differs from the one presented here. In GF, there was a shoulder in
the mass density profile at a radius of 2.0 parsecs (about
0.8\arcmin), which is not present in Fig.~7.  The main reason for the
difference is the increase in the size of our velocity sample by a
factor of two, which obviously reduces the noise in our estimation.  A
related effect is the increased radial coverage of the CFHT data,
which spans more than two decades compared to one for the CTIO data in
GF.  This larger coverage makes it easier to distinguish noise from
trends and so choose a realistic smoothing.  This choice presents more
of a problem with non-parametric techniques that impose smoothing on
the projected quantities, as we do here, rather than ones that impose
smoothing in the space of the desired function (see Merritt \&
Tremblay 1994), since the noise in the projected quantities will be
amplified upon deprojection. We feel that the mass density profile for
M15 in GF may have been undersmoothed.

\subsection{Black Hole Models}

We can also invert the above procedure by assuming a mass density
profile for M15 and calculating the projected velocity dispersion
implied by the isotropic Jeans equation.  Figure~9 shows the projected
dispersion profiles that result from assuming that M15 has a uniform
stellar M/L and a central black hole with various values for the mass.
The dashed and dotted lines are the estimated velocity dispersion
profile and the 90\% confidence band from Fig~6.  The solid lines are
the expected velocity dispersion profiles assuming a stellar M/L of
1.7 and black hole masses of 0, 500, 1000, 3000, and 6000~$\Msun$.
Note that all of these models with an M/L that does not change with
radius are just excluded at 90\% confidence because they rise too
steeply between a radius of 100\arcsec\ and 10\arcsec.  We will
discuss the possibility of a black hole at the center of M15 in \S~5.

\subsection{Mass Functions}

We can estimate the present-day mass function of M15 using the
technique described in GF.  The Jeans equation relates the cluster
potential and the number density and velocity dispersion profiles of a
tracer population.  We calculate the cluster potential from the mass
density determined in \S~4.1.  If we know the velocity dispersion
profile for some population, then we can calculate the number density
profile for that population up to a multiplicative constant. As in GF,
we will assume local thermodynamic equilibrium (LTE); LTE allows us to
relate the observed dispersion profile of the giants and turnoff stars
to the profiles of objects with other masses.  This method will only
yield an approximate result, since LTE is not strictly valid at any
radius and is, moreover, a poor approximation for low-mass objects at
large radii.  The mass density profiles for the individual mass
components must sum to the total density profile. We determine the
necessary multipliers for each number density profile using maximum
penalized likelihood. To keep the derived mass function reasonably
smooth, we use the second derivative of the mass function as the
penalty function (equations 7--9 of GF).

Figure~10 plots the resulting mass functions for M15.  The different
lines represent different radial ranges and their 68\% confidence
bands: the solid lines are the mass function for the inner 25\% of the
mass (0.0--2.0~pc); the dashed lines that for the 25--50\% mass range
(2.0--5.0~pc); and the dotted lines that for the 50--75\% mass range
(5.0--8~pc).  The solid points are located at the masses used in the
fitting.

Both luminous and non-luminous objects contribute to the mass
functions in Fig.~10. The global cluster M/L provides a constraint on
the relative contributions, but exploiting this requires knowing the
uncertain light-to-mass ratios (L/M) for the main-sequence and giant
stars in the various mass groups. We begin by calculating the cluster
M/L assuming that all of the stars with masses below the turn-off are
luminous. Multiplying the L/M's of the various mass groups by the mass
function yields the total cluster luminosity, which we may then divide
into the dynamically-estimated total cluster mass. The L/M values that
we used are based on direct estimates from color-magnitude diagrams
(see Pryor \etal\ 1991). The values are (m is the stellar mass in
solar units) zero for m$<0.16$ and m$>0.8$, 0.014 for $0.16<{\rm
m}<0.25$, 0.026 for $0.25<{\rm m}<0.40$, 0.14 for $0.40<{\rm m}<0.63$,
4.6 for $0.6<{\rm m}<0.8$ (including giant stars). The ratio of this
population M/L to the average dynamical M/L derived from the
normalization of the velocity dispersion profile is approximately the
fractional contribution of the main-sequence and giants stars to the
total mass. The M/L derived from the velocity dispersion normalization
is 1.7 (\S 4.2), and the M/L derived from the mass function is
0.3. The ratio implies that 85\% of mass is in non-luminous objects
not included in the L/M values, presumably stellar remnants.

The mass functions in Fig.~10 argue that much of the mass of M15 is in
the form of 0.6--0.7~$\Msun$ objects, presumably white dwarfs.  GF
reached the same conclusion for M15 and three other clusters. This
result may be in conflict with the mass of 0.5~$\Msun$ found by Richer
\etal\ (1995) for white dwarfs in the globular cluster M4 based on
their location in the color-magnitude diagram. However, some heavier
white dwarfs are also expected from the evolution of massive
main-sequence stars early in the history of the cluster.

The constraints on the numbers of lower-mass objects ($<$ 0.3
M$_\odot$) are not as firm due to the uncertainty in the velocity
dispersion at large radii, as reflected in the larger confidence band,
and due to the assumption of LTE. A more reasonable approach would be
to estimate properly the relation between the velocity dispersion
profiles for different masses using Fokker-Planck
simulations. However, we find relatively few objects with masses below
$0.3~\Msun$. This last result contradicts GF, who found mass functions
increasing at the smallest masses for all of the clusters that they
studied, including M15. The different result yielded by the larger
velocity dataset presented here demonstrates the sensitivity of the
mass function estimate to noise in the data.

\section{SUMMARY AND DISCUSSION}

Knowledge of the present-day stellar mass function in globular
clusters is crucial for understanding the initial mass function.  The
latter contains important information about star formation in the
early galaxy and plays an important role in cluster dynamical
evolution (e.g., Angelletti \& Giannone 1979; Chernoff \& Weinberg
1990).  Dull~\etal\ (1996) have produced an extensive set of
Fokker-Planck models for M15 and find that the projected dispersion
profile is well fitted by a post-collapse model.  From this model,
they estimate that there are a few $\times$ $10^4$ objects more
massive than 1~$\Msun$ in the central 6\arcsec\ (0.3~pc).  Integrating
our estimated mass function (Fig.~10) over objects more massive than
1~$\Msun$ gives about 2000 objects, which are all in the innermost
25\% of the mass (radii smaller than 2.0~pc).  Both models find a few
$\times$ $10^5$ objects with masses of about 0.7~$\Msun$, which are
presumably a combination of main-sequence stars and stellar remnants
(white dwarfs). The largest disagreement between their and our mass
function is for the numbers of low-mass objects.  Dull~\etal\ find a
few $\times 10^5$ stars with masses less than 0.3$\Msun$, which is
higher than our 95\% confidence level for these masses.

It needs to be stressed that both of the above dynamical estimates of
the present-day mass function are uncertain: our results will be
inherently noisy since we are trying to estimate the quantities
directly from the data, while Dull~\etal\ require assumptions about
the initial mass function and the relation between the initial stellar
mass and the final remnant mass, both of which may have large
uncertainties.  The strongest conclusions about the M15 mass function
are those found by both: that a large number (2000 to $10^4$) of
1.4~$\Msun$ objects are present in the central region and that most of
the cluster mass is in 0.5-0.7~$\Msun$ objects.  These conclusions can
be compared to those of Heggie \& Hut (1996), who find that
approximately 50\% of 47~Tuc's mass may be in the form of white dwarfs
or lower-mass stars.  Our results suggest that white dwarfs, rather
than lower-mass stars, constitute most (about 85\%) of the mass of
M15.

Strengthening our dynamical estimates of M15's mass function requires
larger number of velocities in the outer regions ($R>2$\arcmin) of
M15.  Low-mass stars are expected to populate the regions at large
radii, where photometric studies have shown that there may be
significant numbers of low-mass stars (Hesser \etal\ 1987, Richer
\etal\ 1990).  Since photometric estimates of the number of low-mass
stars require large completeness corrections and depend on the
uncertain main-sequence mass-luminosity relation, dynamical
estimates provide valuable additional constraints.

The theory of core collapse appears to pass a very basic test: the
mass density profile for M15 shown in Fig.~7 is remarkably well fit by
a power law with an index of --2.2 throughout most of its radial
extent, as predicted (Cohn 1980).  There is a shallow ``shoulder'' in
the profile at a radius of about 0.2\arcmin\ which is suggestive of a
transition between most of the density being provided by massive
stellar remnants and most by stars with the turnoff mass (the latter
having a shallower profile).  Unfortunately, Dull \etal\ (1996) did
not show the spatial density profiles of their Fokker-Planck models.
The projected density profiles (their Fig.~10) do suggest that this
transition should occur at about that radius.  However, our measured
projected velocity dispersion profile in Fig.~9 is somewhat flatter
than the profiles predicted by the the Dull \etal\ models (see their
Fig.~6), which increase from a value of 9.5~\kms\ at 30\arcsec\ to
12~\kms\ at 3\arcsec.

Rotation has generally not been included in calculations of globular
cluster evolution and we stress the importance doing so in light of
our results for M15.  The radial profiles of the rotation PA and
amplitude, such as are shown in Fig.~5, could provide significant
constraints for such models.  Recently, Einsel \& Spurzem (1996) have
produced Fokker-Planck simulations including rotation.  Their
projected rotation curve is similar to what we measure using the
stellar velocities, presented as the solid points in Fig.~5: there is
a peak in the rotation amplitude close to the half-mass radius.
Unfortunately, we have too few velocities at large radii to allow a
detailed comparison there to the results of Einsel \& Spurzem.  Unlike
the broad agreement at larger radii, in the central region we find a
possible increase in the rotation amplitude which is not seen in their
results.  One explanation could be the existence of a central mass
concentration not included in the models.  Since the models contain
only a single-mass stellar component, possibilities are both heavy
remnants and a massive central black hole.  We also see a large change
in the rotation position angle at small radii, which is even harder to
understand theoretically.  Both the theory and the data need to be
improved for the region near the center of the cluster.

Fig.~9 suggests that the maximum allowed black hole mass is around
3000~$\Msun$. However, this number is uncertain due to our assumptions
of isotropy and constant M/L. Fully 2-integral, and possibly
3-integral, models are necessary to adequately constrain the maximum
allowed black hole mass from the velocity dispersion data alone.

Whether or not M15 contains a central black hole remains open. Fig.~9
suggests that the best isotropic model with a constant stellar M/L is
one that contains a black hole of 1000~$\Msun$. However, if the M/L is
allowed to vary, then the velocity dispersion data imply a change from
1.7 at large radii to only 3 in the central region (Fig. 7).  This
barely significant change could be caused by only the modest number of
heavy-remnants predicted in the central region of old globular
clusters by Fokker-Planck simulations such as those of Dull~\etal\
(1996).  Both models -- a 1000 $\Msun$ central black hole and an
increase of the M/L to 3 -- have the same projected velocity
dispersion profile.  Thus, even with additional velocity data in the
central regions, it will be difficult to discriminate between
them. One alternative is to use the rotation properties to
discriminate between the models. For instance, a central mass
concentration could cause an increase in the rotation amplitude at
smaller radii, as has been observed in Fig.~5.  Determining the
strength of such a test will require detailed modeling including a
central mass concentration and more sophisticated Fokker-Planck
simulations.

In summary, even excellent ground-based seeing ($\sim0.8\arcsec$)
still imposes significant uncertainties on the rotation and velocity
dispersion profiles of M15 at small radii.  Better data would reduce
the present uncertainties in these profiles in the central region and
might reveal behavior that cannot be produced by the change in M/L
resulting from mass segregation.  For example, the velocity dispersion
produced by a 3000~$\Msun$ central black hole would lead to an
increase of the inferred M/L to around 8 at a radius of 0.03~pc, which
would be difficult to obtain with any reasonable remnant population.
The data to test such predictions will require adaptive optics or HST.

\acknowledgments

KG would like to thank the Sigma $\chi$ foundation, which provided
support for travel and accommodations during the observations.  We
also thank Jennifer Gieber for reducing the radial velocity standard
star data and the staff at the CFHT who made two very difficult runs
work smoothly.  Partial support for this research was provided by
grant AST90-20685 from the National Science Foundation.

\clearpage
\begin{figure}
\plotfiddle{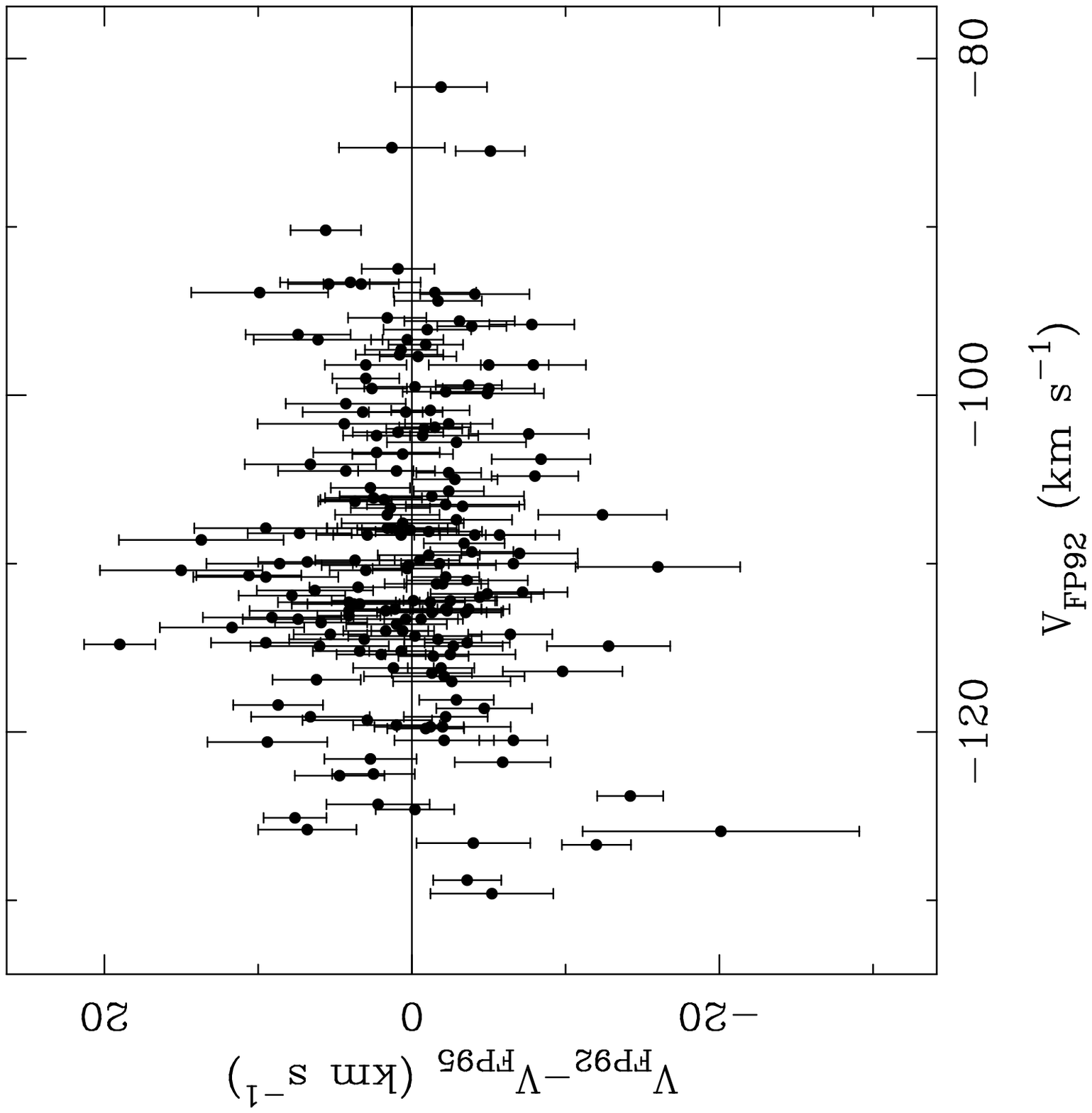}{300pt}{-90}{66}{66}{-260pt}{425pt}
\caption{Comparison of the Fabry-Perot velocities for stars in M15 with
the Fabry-Perot velocities of Paper~1.}
\end{figure}

\clearpage
\begin{figure}
\plotfiddle{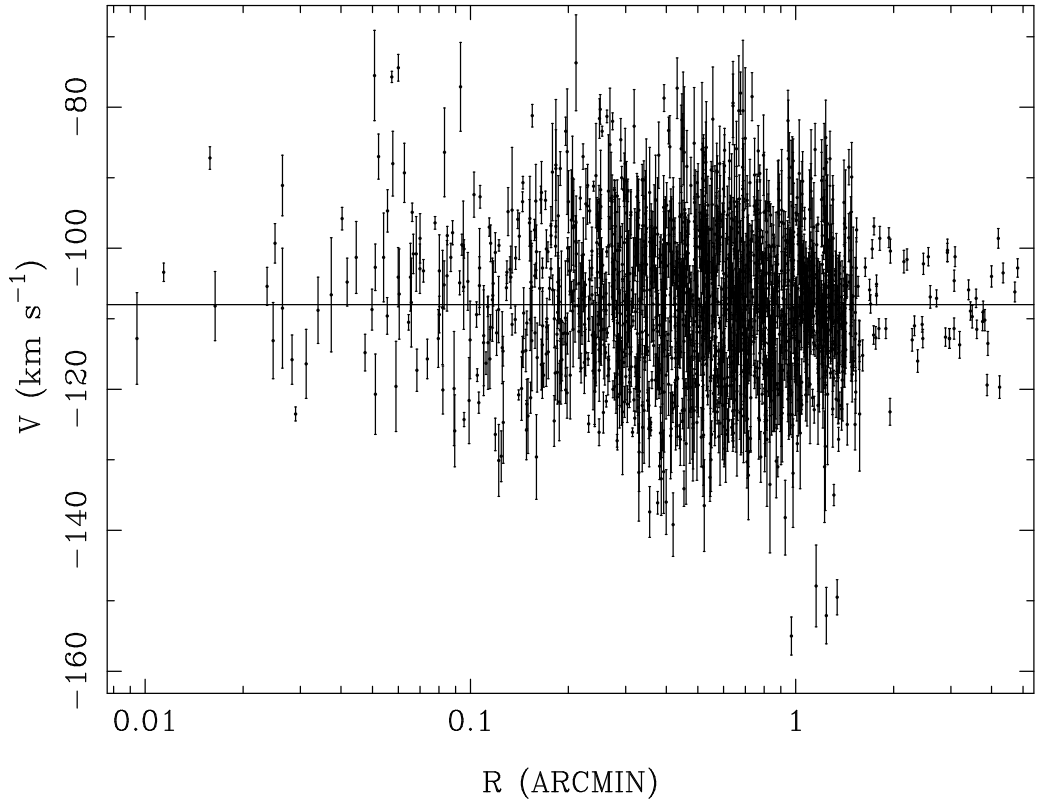}{300pt}{-90}{66}{66}{-260pt}{425pt}
\caption{Radial velocities plotted vs. radius from the center of
M15. The solid horizontal line is at the sample mean of
$-107.8\pm0.3$~\kms.}
\end{figure}

\clearpage
\begin{figure}
\plotfiddle{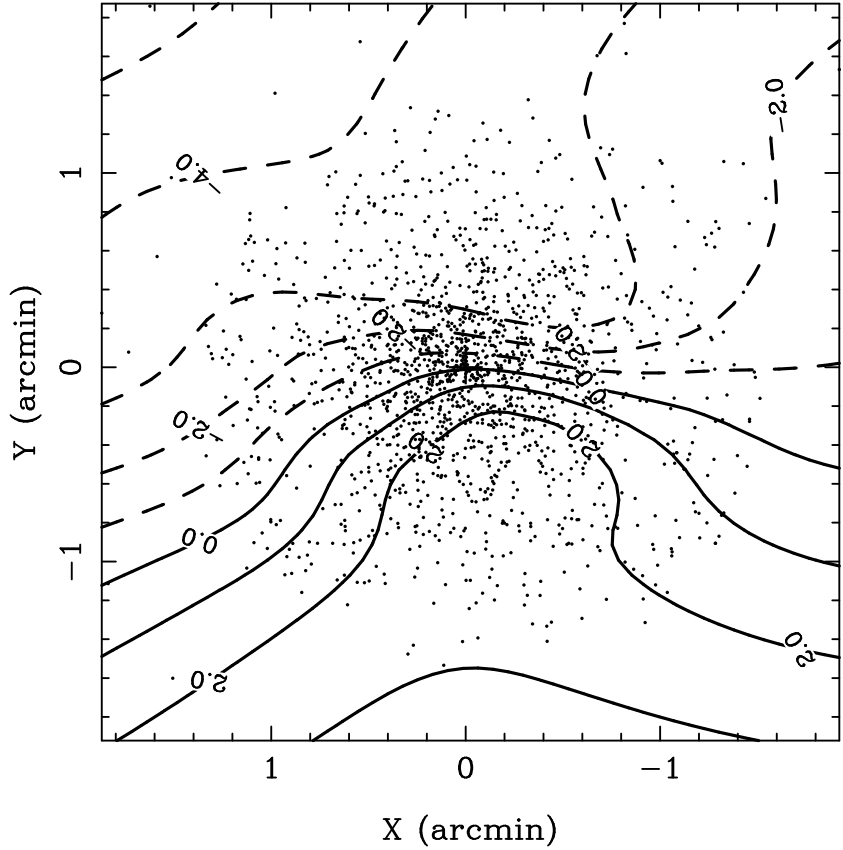}{300pt}{-90}{66}{66}{-260pt}{425pt}
\caption{Velocity map of M15 using a thin-plate smoothing spline
applied to the individual stellar velocities. The cross is at the
center of the cluster. The points represent the positions of the star
with measured velocities. The contours are spaced by 1~\kms, with
dashed lines representing negative contours.}
\end{figure}

\clearpage
\begin{figure}
\plotfiddle{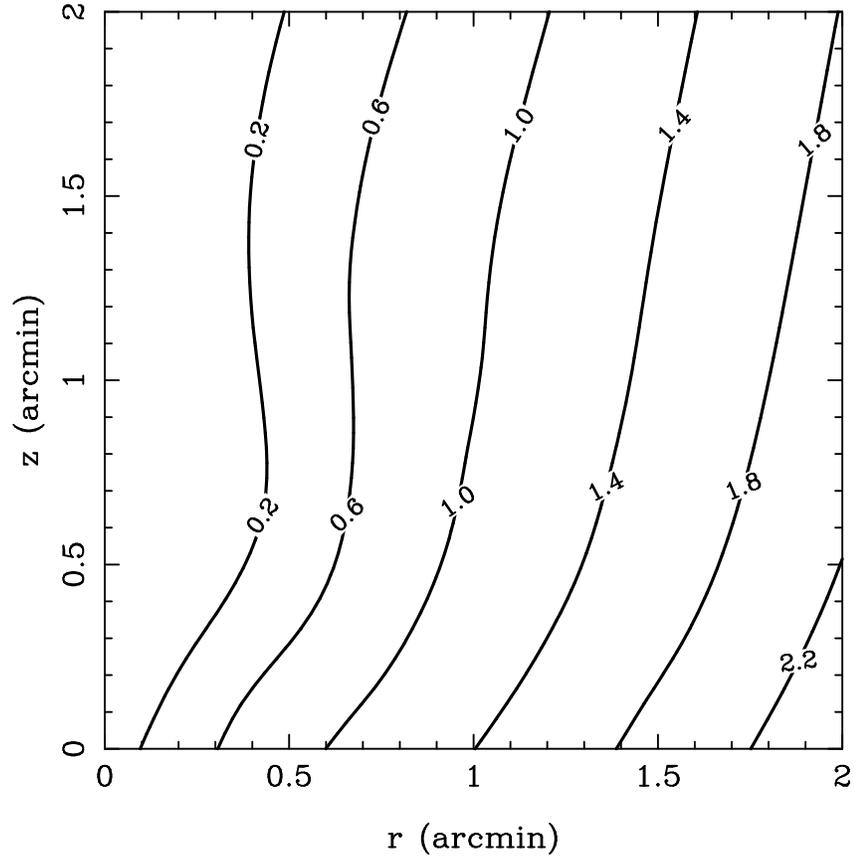}{300pt}{-90}{66}{66}{-260pt}{425pt}
\caption{Rotation of M15 obtained with a thin-plate smoothing spline.
The r and z axis are perpendicular and along the rotation axis
(220\deg) of the cluster, respectively. The contours are in \kms.}
\end{figure}

\clearpage
\begin{figure}
\plotfiddle{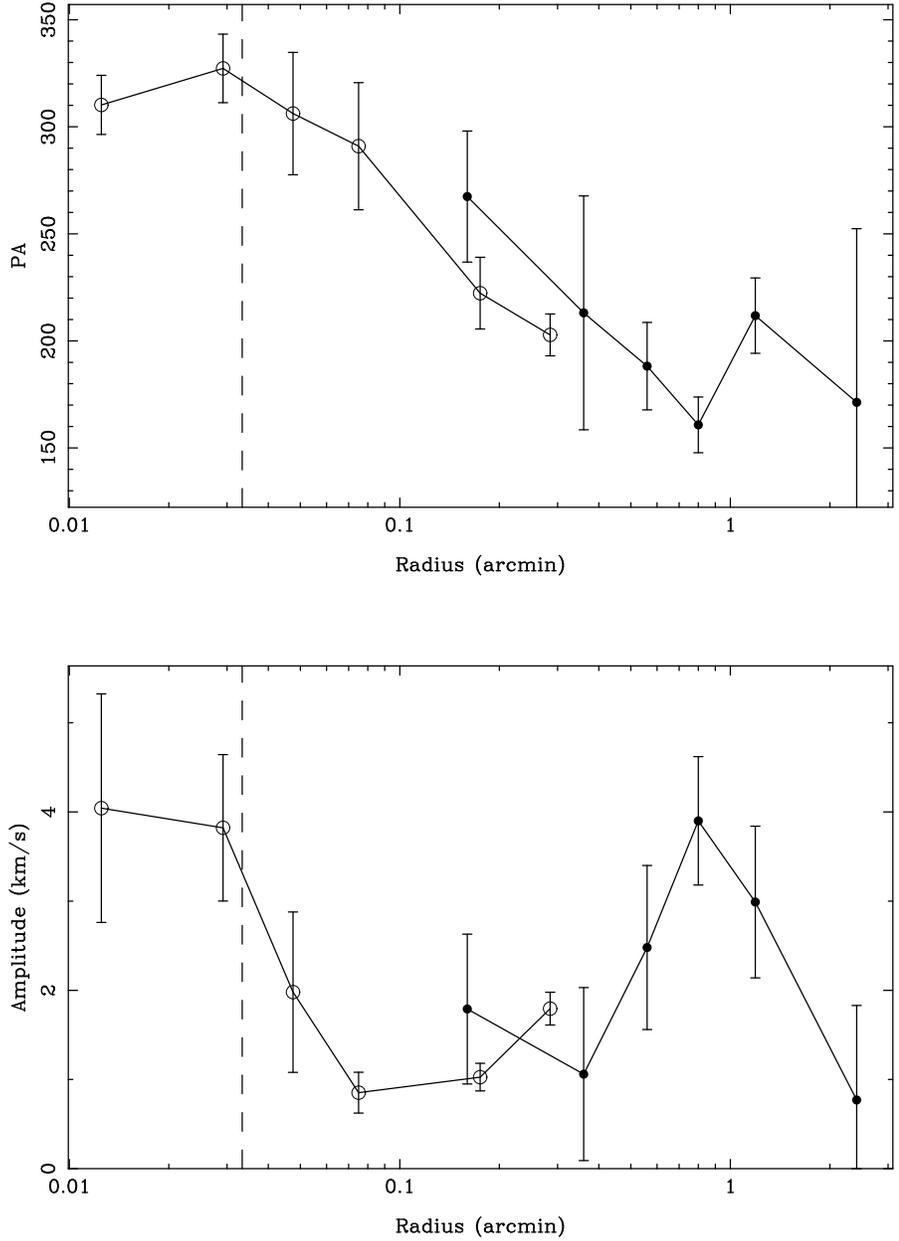}{400pt}{0}{66}{66}{-210pt}{-20pt}
\caption{Rotation position angle and amplitude as a function of
radius. The solid points come from the best fit sinusoid to the
individual velocities in radial bins of 300 stars. The open circles
are the best-fit parameters for the azimuthally binned and averaged
velocities from the integrated-light velocity map. The dashed line is
the estimate of the radius (2\arcsec) where seeing may significantly
affect the estimate of the rotation from the integrated light.}
\end{figure}

\clearpage
\begin{figure}
\plotfiddle{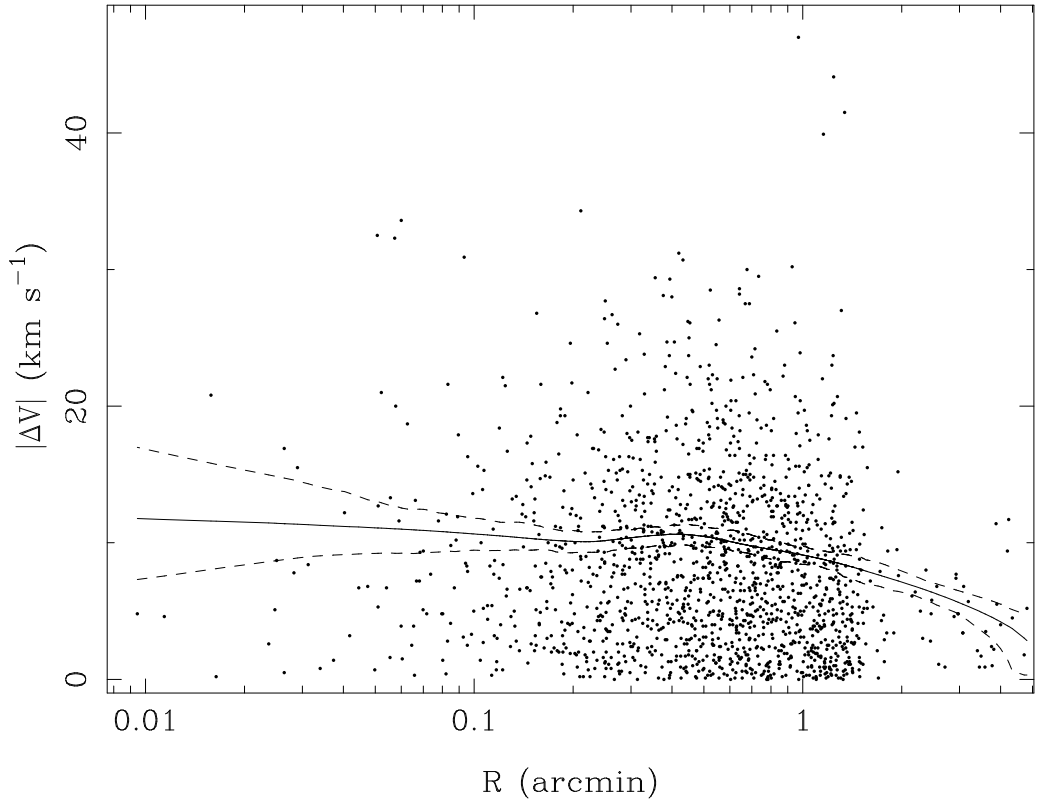}{300pt}{-90}{66}{66}{-260pt}{425pt}
\caption{Velocity dispersion of M15. The points are the absolute
deviations of the individual velocity measurements from the cluster
mean velocity ($-107.8\pm0.3$~\kms). The solid line represents the
LOWESS estimate of the velocity dispersion, and the dashed lines are
the 90\% confidence band. For M15, 1\arcmin\ is 2.8 parsecs.}
\end{figure}

\clearpage
\begin{figure}
\plotfiddle{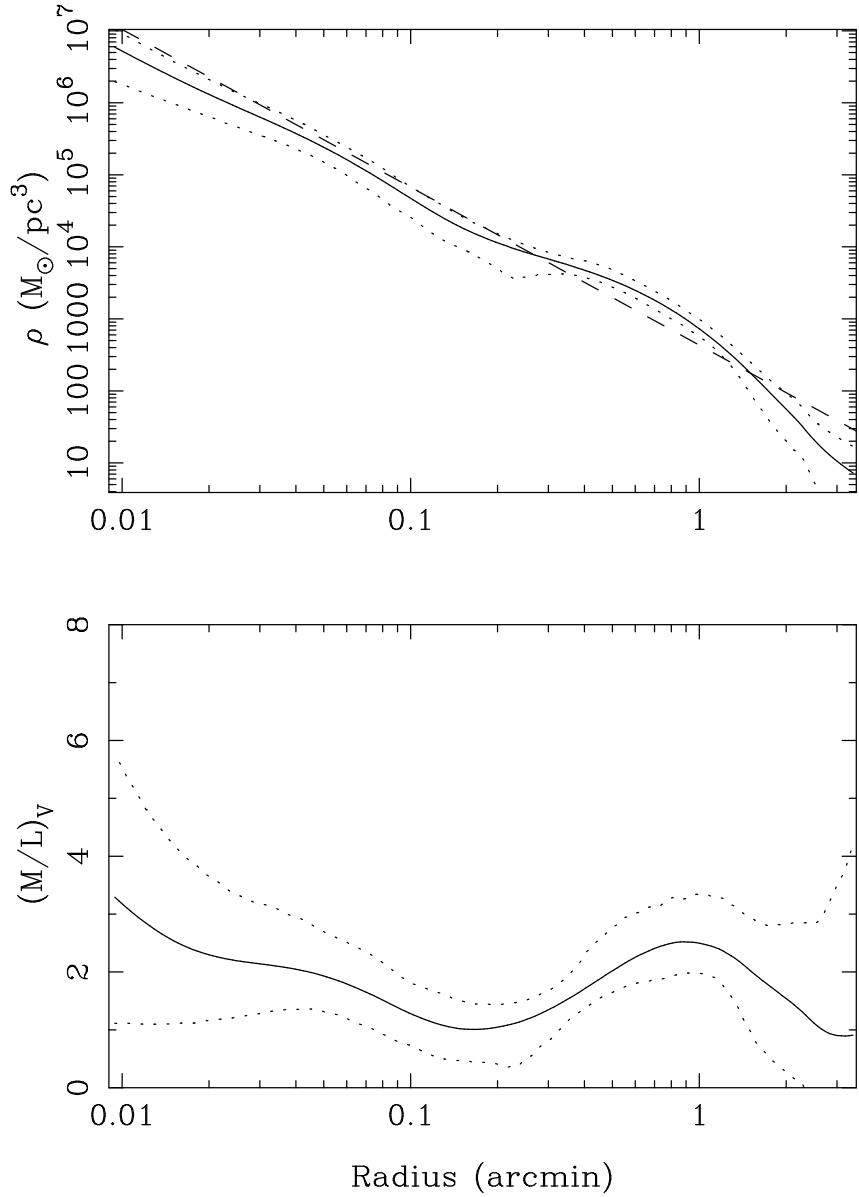}{400pt}{0}{66}{66}{-210pt}{-20pt}
\caption{Mass density and M/L ratio profile for M15. The solid lines
are the non-parametric estimates and the dotted lines are the 90\%
confidence bands. The dashed line in the upper figure is a power-law
slope of --2.23.}
\end{figure}

\clearpage
\begin{figure}
\plotfiddle{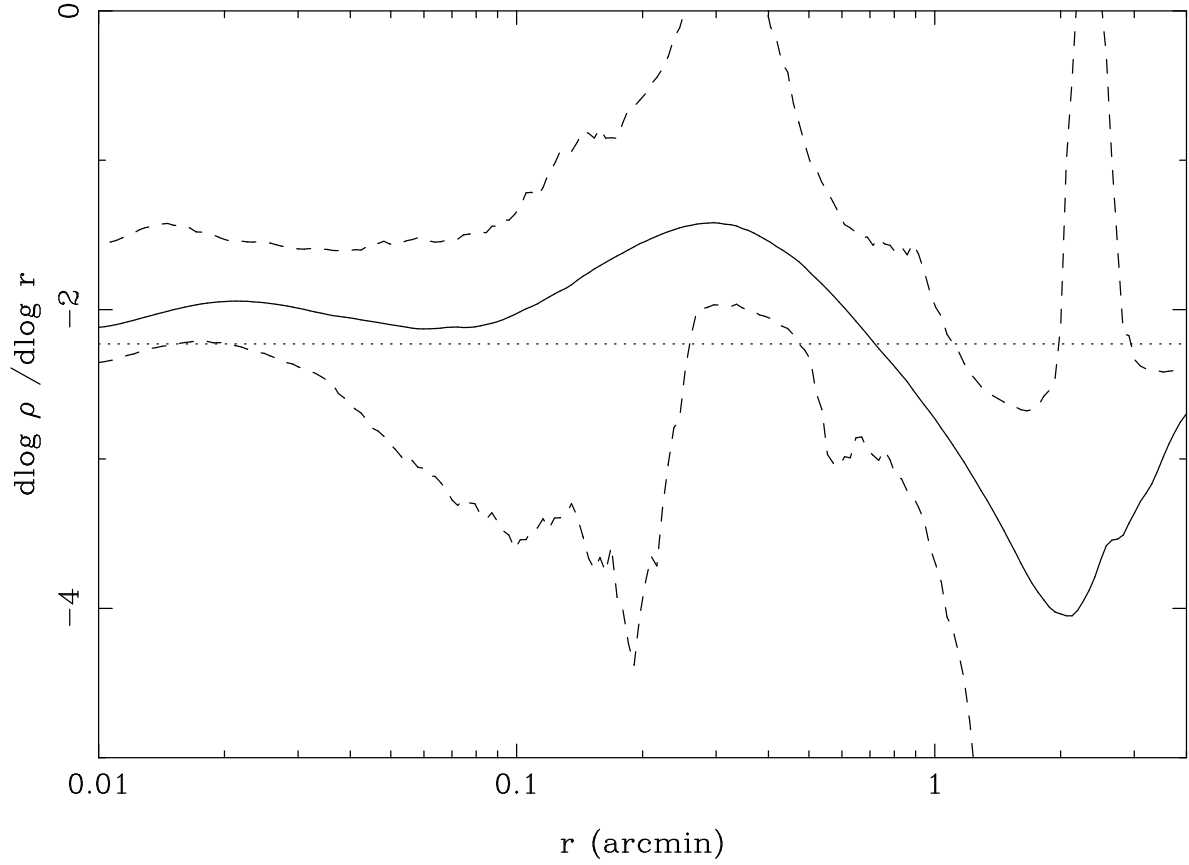}{300pt}{-90}{66}{66}{-260pt}{425pt}
\caption{Logarithmic slope of the mass density as a function of
radius. The solid and the dashed lines are the slope and the 90\%
confidence bands, respectively. The dotted line is at the value
--2.23.}
\end{figure}

\clearpage
\begin{figure}
\plotfiddle{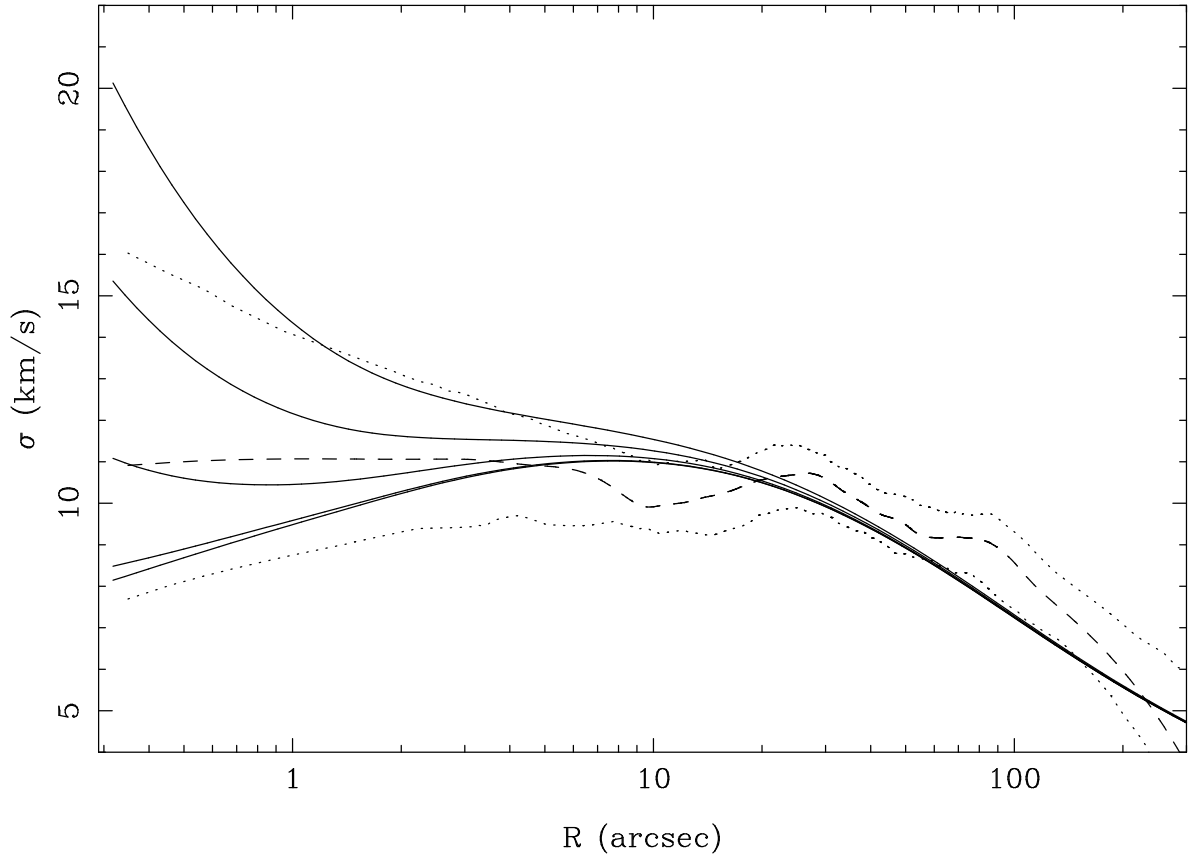}{300pt}{-90}{66}{66}{-260pt}{425pt}
\caption{The dashed and dotted lines are the estimate of the velocity
dispersion and its 90\% confidence band as a function of radius,
respectively. The solid lines are the projected velocity dispersion
assuming constant M/L=1.7, isotropy, and central black hole masses of
0, 100, 1000, 3000, and 6000~$\Msun$.}
\end{figure}

\clearpage
\begin{figure}
\plotfiddle{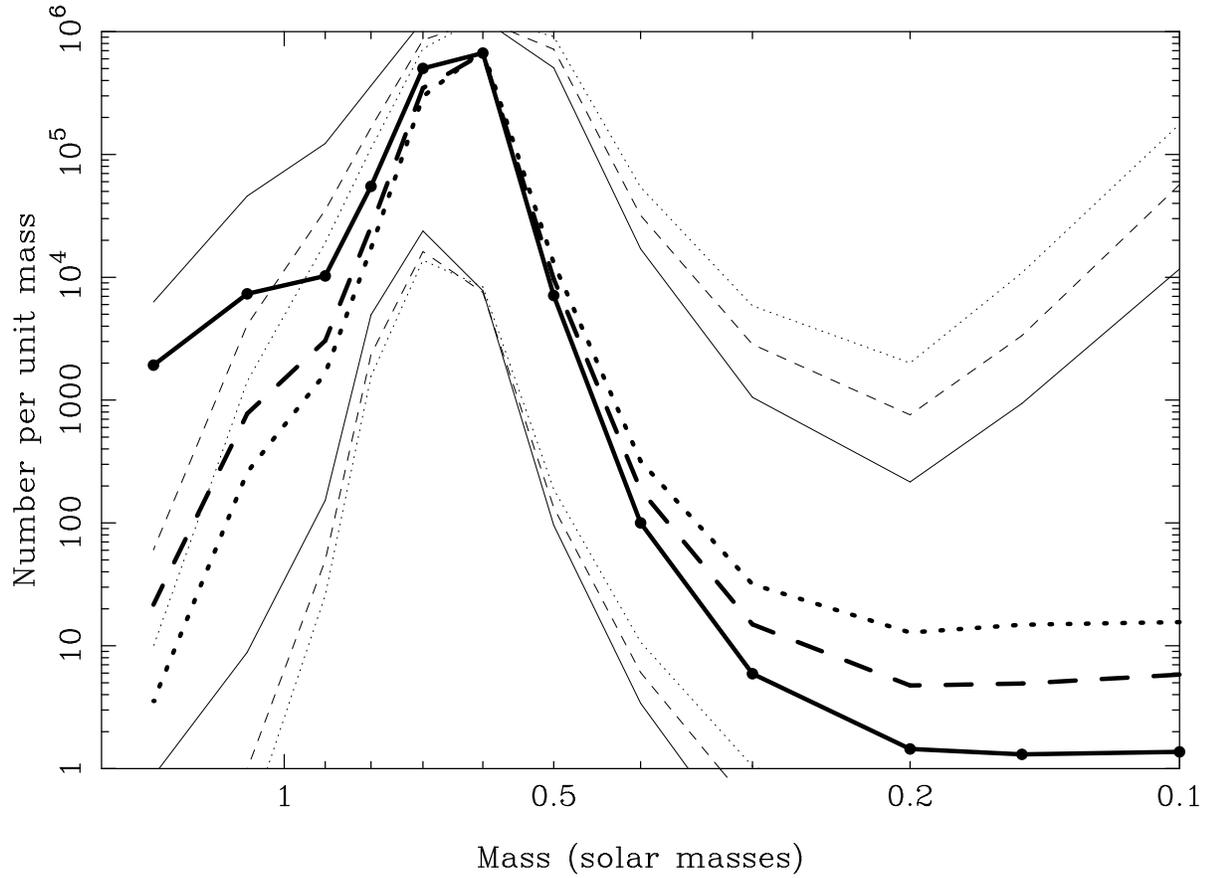}{300pt}{-90}{66}{66}{-260pt}{425pt}
\caption{Mass function for M15. The heavy and light solid lines
represent the values for the inner 25\% mass and its 90\% confidence
band, respectively. The heavy dashed and dotted lines are for the
second and third quartiles of the mass, respectively, and the
corresponding light lines are their confidence bands.}
\end{figure}

\end{document}